\documentclass[11pt]{article}
\usepackage[affil-it]{authblk}
\usepackage{geometry}               
\usepackage{graphicx}
\usepackage{amssymb}
\usepackage{amsmath}
\usepackage{epstopdf}
\usepackage{fullpage}
\usepackage{cite}
\usepackage{bm}
\usepackage{cancel}

\title{Deriving the exact nonadiabatic quantum propagator in the mapping variable representation}
\author{Timothy J.\ H.\ Hele\thanks{On intermission from: Jesus College, University of Cambridge, UK.}
\ and Nandini Ananth\thanks{Email: ananth@cornell.edu}}
\affil{Department of Chemistry and Chemical Biology, Cornell University, Ithaca, New York 14853, USA}

\newcommand{\CABNt}{C_{AB}^{[N]}(t)}

\newcommand{\eb}{e^{-\beta \hat H}}

\newcommand{\ebN}{e^{-\beta_N \hat H}}

\newcommand{\etf}{e^{-i \hat H t/\hbar}}
\newcommand{\etb}{e^{i \hat H t/\hbar}}

\newcommand{\bra}[1]{\langle #1 |}
\newcommand{\ket}[1]{| #1 \rangle}
\newcommand{\kb}[1]{\ket{#1}\bra{#1}}
\newcommand{\bk}[2]{\bra{#1} #2 \rangle}

\newcommand{\dd}[2]{\frac{d #1}{d #2}}

\newcommand{\ddp}[2]{\frac{\partial #1}{\partial #2}}

\newcommand{\piNz}{\prod_{i=0}^{N-1}}

\newcommand{\pinjNz}{\prod_{i=0,\ i\neq j}^{N-1}}

\newcommand{\smiNz}{\sum_{i=0}^{N-1}}

\newcommand{\smjNz}{\sum_{j=0}^{N-1}}
\newcommand{\smkNz}{\sum_{k=0}^{N-1}}

\newcommand{\lNti}{\lim_{N\to \infty}}

\newcommand{\largeN}{$N\to\infty$}

\newcommand{\tr}{ {\rm Tr} }

\newcommand{\no}{\nonumber}

\newcommand{\betaN}{\beta_N}
\newcommand{\ola}{\overleftarrow}
\newcommand{\ora}{\overrightarrow}

\newcommand{\T}{^{\mathrm{T}}}

\newcommand{\re}{\mathrm{e}}
\newcommand{\rW}{\mathrm{W}}
\newcommand{\bp}{ {\bf p} }

\newcommand{\bq}{ {\bf q} }

\newcommand{\bDelta}{ \bm{\Delta} }

\newcommand{\bP}{ {\bf P} }
\newcommand{\bD}{ {\bf D} }
\newcommand{\bR}{ {\bf R}}

\newcommand{\bV}{\mathbf{V}}

\newcommand{\bC}{\mathbf{C}}

\newcommand{\mK}{\mathcal{K}}
\newcommand{\mS}{\mathcal{S}}

\newcommand{\eqr}[1]{Eq.~\eqref{eq:#1}}

\newcommand{\eql}[1]{\label{eq:#1}}
\newcommand{\figr}[1]{Fig.~\ref{fig:#1}}
\newcommand{\figl}[1]{\label{fig:#1}}

\newcommand{\mL}{\mathcal{L}}

\begin{document}
\newcommand{\cev}[1]{\reflectbox{\ensuremath{\vec{\reflectbox{\ensuremath{#1}}}}}}

\bibliographystyle{tim}
\maketitle

\begin{abstract}
We derive an exact quantum propagator for nonadiabatic dynamics in multi-state systems using the mapping variable representation, where classical-like Cartesian variables are used to represent both continuous nuclear degrees of freedom and discrete electronic states. The resulting expression is a Moyal series that, when suitably approximated, can allow for the use of classical dynamics to efficiently model large systems. We demonstrate that different truncations of the exact propagator lead to existing approximate semiclassical and mixed quantum-classical methods and we derive an associated error term for each method. Furthermore, by combining the imaginary-time path-integral representation of the Boltzmann operator with the exact propagator, we obtain an analytic expression for thermal quantum real-time correlation functions. These results provide a rigorous theoretical foundation for the development of accurate and efficient classical-like dynamics to compute observables such as electron transfer reaction rates in complex quantized systems. 
\emph{Submitted to the \emph{Reaction Rate Theory} Faraday Discussion on 2nd May 2016.}
\end{abstract}

\section{Introduction}
The accurate calculation of nonadiabatic dynamics has been a longstanding problem in chemical physics since the 1930s \cite{lan32a,zen32a}, being fundamental to charge and energy transfer in biological and chemical systems \cite{mar85a,wor04a}.  Many approximate methods have been developed using classical, or classical-like dynamics to describe nonadiabatic quantum processes with the electronic degrees of freedom treated as discrete states, including Marcus theory \cite{mar85a,mar65a,mar64a}, surface hopping \cite{tul90a,tul12a,sub10a,sub11a}, semiclassical \cite{sun97a} and mixed quantum-classical \cite{pfa15a,rya14a,wal16a} methods. 

A particularly successful approach involves the use of mapping variables, where discrete electronic degrees of freedom are mapped onto continuous positions and momenta of fictitious harmonic oscillators. Originally proposed by Meyer and Miller \cite{mey79a,mey79b}, this mapping was shown to be exact by Stock and Thoss \cite{sto97a,sto05a} and has been developed using various semiclassical \cite{ana07a,sun98a,rab99a}, quasiclassical \cite{cot16a}, (partially) linearized \cite{huo11a,huo12a,kim08a,kap06a,hsi12a,hsi13a,bon05a,bon05b}, and path integral\cite{ric13a,ana13a,duk15a,ana10a} techniques.

Here, we derive from first principles the exact nonadiabatic quantum propagator in the mapping variable representation and relate this to the conventional adiabatic (single surface) propagator, the Moyal series \cite{moy49a,gro46a}. We show that careful approximation of the exact propagator leads to a number of existing methods, and we provide the associated error term in each case. Furthermore, using the generalized Kubo transform \cite{hel13a,alt13a,hel13b}, previously employed to obtain approximate quantum dynamics methods in single-surface systems \cite{hel15a,hel15b,hel16a}, we obtain an analytic expression for the nonadiabatic quantum Boltzmann distribution and the exact propagator in the path-integral representation. 

The article is structured as follows: in section~\ref{sec:bg}, we provide an overview of background theory, in section~\ref{sec:cor} we derive the exact propagator and make approximations that lead to various existing methods. Thermal correlation functions are discussed in section~\ref{sec:therm}, in section~\ref{sec:multin} we obtain an exact path-integral propagotar using Generalized Kubo transform formulation, and we present our conclusions in section~\ref{sec:con}.

\section{Background theory}
\label{sec:bg}
The background theory for mapping variables and Wigner transforms are reviewed here to provide context for the main body of the article; for a detailed review of mapping variables and other nonadiabatic techniques, see Ref~\citenum{sto05a}. 

\subsection{Mapping variables}
For simplicity we consider a system with one Cartesian dimension position $R$ with conjugate momentum $P$, mass $m$ and $\mK$ diabatic electronic states with overall Hamiltonian\footnote{We also assume a sufficiently high temperature that exchange effects may be neglected.}
\begin{align}
\hat H = & \frac{\hat P^2}{2m} + V_0(\hat R) + \sum_{n,m=1}^{\mK} \ket{\phi_n} V_{\rm e}(\hat R)_{nm} \bra{\phi_m} \eql{ham} \\
\equiv & \frac{\hat P^2}{2m} + \sum_{n,m=1}^{\mK} \ket{\phi_n} [V_{\rm e}(\hat R)_{nm} + \delta_{nm}V_0(\hat R)] \bra{\phi_m} \eql{ham2},
\end{align}
where $V_0(\hat R)$ is the state-independent part of the potential and $V_{\rm e}(\hat R)_{nm}$ is a matrix element of the nonadiabatic potential matrix $\bV_{\rm e}(\hat R)$. The equivalence of \eqr{ham} and \eqr{ham2} follows from application of the identity 
\begin{align}
\mathbf{\hat I} = \int dR \sum_{n=1}^{\mK} \kb{R,\phi_n} \eql{id}.
\end{align}
We assume throughout that $\bV_\re (\hat R)$ is real and symmetric; extension to a complex hermitian Hamiltonian (and multidimensional systems) is straightforward.

The Hamiltonian can equivalently be written in the singly excited oscillator (SEO) basis $\{\ket{n}\}$, $n=1,\ldots,\mK$, where $\ket{n}$ corresponds to one quantum of excitation in the $n$th oscillator and zero quanta in the remaining $\mK-1$ oscillators\cite{mey79a,mey79b,sto97a,sto05a}. This is achieved by mapping
\begin{align}
\ket{\phi_n}\bra{\phi_m} \to \hat a^\dag_n \hat a_m,
\end{align}
where $\hat a^\dag_n$ creates one quantum of excitation in the $n$th oscillator and $\hat a_m$ destroys one quantum in the $m$th oscillator. 
An Operator $\hat O$ in the diabatic representation can then be expressed as
\begin{align}
\hat O = \sum_{n,m=1}^{\mK} \hat a^\dag_n O(\hat R, \hat P)_{nm} \hat a_m
\end{align}
where $O(\hat R,\hat P)_{nm}$ is a scalar [matrix element of $\mathbf{O}(\hat R, \hat P)$] in the space of electronic states, but an operator in the space of nuclear co-ordinates and momenta. 

Writing the creation and annihilation operators in the position and momentum representation 
\footnote{following others, we set the (arbitrary) mass and frequency of the harmonic oscillators to unity in atomic units but 
	retain $\hbar$ necessary to construct semiclassical approximations to the propagator.},
\begin{align}
\hat a_m = \frac{1}{\sqrt{2\hbar}} (\hat q_m + i \hat p_m), \qquad \hat a^\dag_n = \frac{1}{\sqrt{2\hbar}} (\hat q_m - i \hat p_m),
\end{align}
we find
\begin{align}
\hat O = & \frac{1}{2\hbar} \sum_{n,m=1}^{\mK} O_{nm}(\hat R, \hat P)(\hat q_n \hat q_m + \hat p_n \hat p_m - \delta_{nm} \hbar). \eql{op}
\end{align}
The only operators in the mapping variable representation which correspond to a physically observable quantity are those of the functional form in \eqr{op}, whose application upon a SEO will stay in the subspace of SEOs \cite{sto97a}. 

The SEO eigenstates in the position representation are
\begin{align}
\bk{\bq}{n} = \sqrt{\frac{2}{\hbar}} \frac{1}{(\pi\hbar)^{\mK/4}} q_n e^{-\bq\cdot\bq/2\hbar}
\end{align}
and the corresponding identity to \eqr{id}
\begin{align}
\mathbf{\hat I} = \int dR \sum_{n=1}^{\mK} \kb{R,n}. \eql{idmv}
\end{align}
The identity expressed in electronic position-space variables,
\begin{align}
\mathbf{\hat I}' = \int dR \int d\bq \ \kb{R,\bq} \eql{idp},
\end{align}
is overcomplete, since it includes all possible excitations of any of the $\mK$ oscillators, rather than just SEO states. However, using the SEO projection operator, $\mathcal{\hat S} = \sum_{n=1}^{\mK} \kb{n}$, we can constrain the position-space identity in \eqr{idp} to the subspace of SEO\cite{ana13a},
\begin{align}
\mathbf{\hat I} = &  \int dR \int d\bq \ \mathcal{\hat S} \kb{R,\bq} \eql{id1} \\
= & \int dR \int d\bq \ \kb{R,\bq}\mathcal{\hat S} \eql{id2} .
\end{align}

\subsection{Wigner transformed operators}
Here we present standard results for Wigner distributions \cite{wig32a,moy49a,hil84a} adapted to the mapping variable representation introduced in the previous section. 

The Wigner transform \cite{wig32a} of an operator in the mapping variable representation is
\begin{align}
[\hat O]_W(R,P,\bq,\bp) = & \int dD \int d\bDelta \ e^{iPD/\hbar} e^{i\bp\cdot\bDelta/\hbar} \no\\
& \times \bra{R-D/2,\bq-\bDelta/2}  \hat O \ket{R+D/2,\bq+\bDelta/2}. \eql{owig}
\end{align}
Inserting \eqr{op} into \eqr{owig} and evaluating the integrals over $\bDelta$ gives 
\begin{align}
[\hat O]_W(R,P,\bq,\bp) = & \frac{1}{2\hbar} \int dD e^{iPD/\hbar} \tr[(\bC - \hbar \mathbf{1}) \bra{R-D/2} \mathbf{O}(\hat R, \hat P) \ket{R-D/2} ] \eql{oeval}
\end{align}
where $\mathbf{1}$ is the $\mK \times \mK$ identity matrix,
\begin{align}
\mathbf{C} = & (\bq + i\bp)\otimes(\bq - i \bp)\T,
\end{align}
and $\hat O$ is written in the matrix representation
\begin{align}
\bra{R-D/2} \mathbf{O}(\hat R, \hat P) \ket{R+D/2}_{nm} \equiv & \bra{R-D/2} O (\hat R, \hat P)_{nm} \ket{R+D/2}. \eql{matrep}
\end{align}
If the projection operator $\hat \mS$ is inserted alongside the operator, the Wigner transform (denoted with a subscript $\mS$) is \cite{ana13a}
\begin{align}
[\hat O_\mS]_W&(R,P,\bq,\bp) \no \\
 \equiv & [ \hat \mS \hat O \hat \mS]_\rW (R,P,\bq,\bp)  \\
= & \int dD \int d\bDelta \sum_{n,m=1}^{\mK}\ e^{iPD/\hbar} e^{i\bp\cdot\bDelta/\hbar} \no\\
& \times \bk{\bq-\bDelta/2}{n} \bra{R-D/2} O (\hat R, \hat P)_{nm} \ket{R+D/2} \bk{m}{\bq+\bDelta/2} \eql{moven} \\
= & \frac{2^{\mK+1}}{\hbar} e^{-G/\hbar} \int dD \ e^{iPD/\hbar} \tr \left[ (\mathbf{C} - \frac{\hbar}{2}\mathbf{1})  \bra{R-D/2} \mathbf{\hat O} (\hat R, \hat P) \ket{R+D/2} \right], \eql{oeval2}
\end{align}
where 
\begin{align}
G= & \bq\cdot\bq + \bp\cdot\bp 
\end{align}
and we have noted that $\ket{n}$ does not depend on $R$ to obtain \eqr{moven}.

In some circumstances $\hat O$ can be written as an nuclear-only part $\hat O_{\rm n}$ and an electronic part $\hat O_{\rm e}$ [such as the Hamiltonian in \eqr{ham}] and the operator in the mapping variable representation becomes
\begin{align}
[\hat O]_\rW(R,P,\bq,\bp) = & [\hat O_{\rm n}]_W + \frac{1}{2\hbar} \tr\left[(\bC - \hbar \mathbf{1}) [\mathbf{O}_{\rm e}(\hat R, \hat P)]_\rW \right] \eql{nucel}
\end{align}
where the nuclear-only Wigner transform is
\begin{align}
 [\hat O_{\rm n}]_\rW = \int dD\ e^{iPD/\hbar} \bra{R-D/2} O_{\rm n}(\hat R, \hat P) \ket{R+D/2}
\end{align}
and the integral over $D$ in \eqr{oeval} has been taken inside the trace of electronic variables
\begin{align}
([\mathbf{O}_{\rm e}(\hat R, \hat P)]_\rW)_{nm} =  \int dD e^{iPD/\hbar} \bra{R-D/2} O(\hat R, \hat P)_{nm} \ket{R+D/2}.
\end{align}

The trace of the product of two operators is simply the integral of the product of their Wigner transforms \cite{hil84a} 
\begin{align}
\tr [\hat O_1 \hat O_2] = \frac{1}{(2\pi\hbar)^{\mK+1}} \int dR \int dP \int d\bq \int d\bp\ [\hat O_1]_\rW [\hat O_2]_\rW,
\end{align}
though in the mapping variable representation this must be combined with SEO identities $(\hat \mS)$ in order to confine the mapping variables to the correct Hilbert space. 
If there is no trace we can adapt the standard relation for the Wigner transform of a product \cite{hil84a,gro46a}
\begin{align}
[\hat O_1 \hat O_2]_\rW = [\hat O_1]_\rW e^{-i\bm{\Lambda} \hbar/2} [\hat O_2]_\rW \eql{wigprod}
\end{align}
to the mapping variable representation, where the $\bm{\Lambda}$-operator is the negative of the Poisson bracket operator
\newcommand{\Lambdan}{\Lambda_{\mathrm{n}}}
\newcommand{\Lambdae}{\bm{\Lambda}_{\mathrm{e}}}
\begin{align}
\bm{\Lambda} = 
\left(\!\!
\begin{array}{c}
\Lambda_\mathrm{n} \\
\bm{\Lambda}_\mathrm{e}
\end{array}
\!\!\right), \eql{lambdadef}
\end{align}
with the nuclear derivative (scalar in one dimension)
\begin{align}
\Lambda_\mathrm{n} = \ola {\partial_P} \ora {\partial_R} - \overleftarrow {\partial_R} \overrightarrow {\partial_P}, 
\end{align}
and the electronic derivative
\begin{align}
\bm{\Lambda}_\mathrm{e} =\ola \nabla_\bp \cdot \ora \nabla_\bq - \ola \nabla_\bq \cdot \ora \nabla_\bp.
\end{align}
We use the shorthand $\partial_P = \ddp{}{P}$ and likewise for $\partial_R$,
\begin{align}
\nabla_\bp = \left(
\begin{array}{c}
\partial_{p_1} \\
\partial_{p_2} \\
\vdots \\
\partial_{p_{\mK}}
\end{array}
\right)
\end{align}
and likewise for $\nabla_{\bq}$, and the arrows represent the direction in which the derivative acts \cite{hil84a, hel15a}.

\section{Correlation functions}
\label{sec:cor}
Consider a general correlation function
\begin{align}
c_{AB}(t) = \tr [\hat A \etb \hat B \etf ] \eql{cor}
\end{align}
whose path-integral form is illustrated in \figr{pidiag}(a). Out of the theoretically infinite possibilities for inserting SEO identities \cite{ric13a,kim08a,ana13a}, we choose the simplest form to construct a Wigner transformed time-evolved operator: we insert \eqr{id1} the right $\hat A$ and \eqr{id2} to the left, followed by Wigner transforming to give
\begin{align}
c_{AB}(t) = \frac{1}{(2\pi\hbar)^{\mK+1}}\int dR \int dP \int d\bq \int d\bp\ [\hat A_\mS]_\rW(R,P,\bq,\bp) [\hat B(t)]_\rW(R,P,\bq,\bp) \eql{cwig},
\end{align}
where we use the shorthand $\hat B(t) = \etb \hat B \etf$. The Wigner-transformed operators $ [\hat A_\mS]_\rW (R,P,\bq,\bp)$ and $[\hat B(t)]_\rW(R,P,\bq,\bp)$ can be evaluated in accordance with \eqr{oeval2} and \eqr{oeval} respectively,
\begin{align}
[\hat A_\mS]_\rW = & \frac{2^{\mK+1}}{\hbar} e^{-G/\hbar} \int dD \ e^{iPD/\hbar} \tr \left[ (\mathbf{C} - \frac{\hbar}{2}\mathbf{1})  \bra{R-D/2} \mathbf{\hat A} \ket{R+D/2} \right], \eql{aeval} \\
[\hat B(t)]_\rW = & \frac{1}{2\hbar} \int dD \ e^{iPD/\hbar} \ \tr\left[(\bC - \hbar\mathbf{1}) \bra{R-D/2} \mathbf{\hat B}(t) \ket{R+D/2} \right], \eql{beval}
\end{align}
where we use the matrix representation of the operators defined in \eqr{matrep} and omit functional dependence of the operators on $(R,P,\bq,\bp)$ (and will continue to do so). The functional form of \eqr{aeval} is slightly more complex than \eqr{beval} due to the presence of SEO identities. 
\begin{figure}[tb]
\centering
\includegraphics[width=0.5\columnwidth]{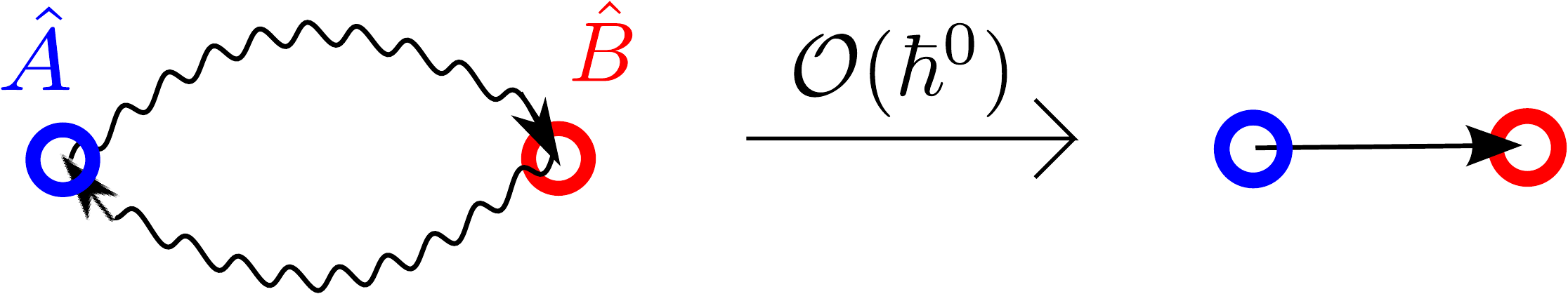}
\caption{Schematic path-integral diagram for the correlation function $c_{AB}(t)$ in \eqr{cor}, showing the effect of truncating the real-time evolution at $\mathcal{O}(\hbar^0)$. Wavy lines represent quantum real-time evolution and straight lines classical time-evolution. Blue and red circles represent $\hat A$ and $\hat B$ respectively, which are assumed to be local.}
\figl{pidiag}
\end{figure}

\subsection{Derivation of the exact propagator}
We now use the Liouvillian formalism \cite{zwa01a,nit06a} to derive an exact propagator in the mapping variable representation. 
Differentiating \eqr{cwig} gives
\begin{align}
\dd{}{t} c_{AB}(t) = \frac{1}{(2\pi\hbar)^{\mK+1}}\int dR \int dP \int d\bq \int d\bp\ [\hat A_\mS]_{\rW} \left[ \frac{i}{\hbar}[\hat H, \hat B(t)]\right]_{\rW}. \eql{difcor}
\end{align}
Using \eqr{wigprod} to expand the Wigner transform of the product of operators in the commutator, we obtain
\begin{align}
\left[ \frac{i}{\hbar}[\hat H, \hat B(t)]\right]_{\rm W}  = \frac{2}{\hbar} [\hat H]_{\rm W} \sin (\bm{\Lambda}\hbar/2) [\hat B(t)]_{\rm W}\eql{hsin}
\end{align}
whose functional form is similar to the Moyal series representation for the adiabatic propagator\cite{hil84a,hel76a}. 
Using \eqr{hsin}, we can now define a Liouvillian\footnote{Following the convention of Zwanzig \cite{zwa01a}, we do not define the Liouvillian with a prefactor of $i$.}
\begin{align}
\mathcal{L} = \frac{2}{\hbar} [\hat H]_{\rm W} \sin (\bm{\Lambda}\hbar/2) \eql{ldef},
\end{align}
and represent the correlation function in \eqr{cwig} as,
\begin{align}
c_{AB}(t) = \frac{1}{(2\pi\hbar)^{\mK+1}}\int dR \int dP \int d\bq \int d\bp\ [\hat A_\mS]_\rW\ e^{\mathcal{L}t} [\hat B(0)]_\rW. \eql{formsol}
\end{align}
To find $\mathcal{L}$ in terms of $\{R, P, \bq, \bp\}$, we first evaluate the Wigner transform of the mapping variable Hamiltonian \cite{sto97a} in \eqr{ham} using \eqr{nucel},
\begin{align}
[\hat H]_{\rm W} = \frac{P^2}{2m} + V_0(R) + V_{\rm e}(R,\bq,\bp) \eql{mvham},
\end{align}
where the nonadiabatic potential has been abbreviated as $V_{\rm e}(R,\bq,\bp)~=~\tr[(\bC-\hbar\mathbf{1})\bV_{\rm e}(R)]/2\hbar$, with $\bV_{\rm e}(R)$ the nonadiabatic potential matrix in \eqr{ham}.
To simplify further, we note that the sine function in \eqr{ldef} can be separated into nuclear and electronic parts,
\begin{align}
\sin(\bm{\Lambda}\hbar/2) = \sin(\Lambdan\hbar/2)\cos(\Lambdae\hbar/2) + \sin(\Lambdae\hbar/2)\cos(\Lambdan\hbar/2). \eql{sinexp}
\end{align}
Since the mapping variable Hamiltonian only contains terms up to second order in $\bp$ and $\bq$, we can without approximation truncate the trigonometric series in $\Lambdae$  to give
\begin{align}
\frac{2}{\hbar} [\hat H]_{\rm W} \sin (\bm{\Lambda}\hbar/2) = &  [\hat H]_{\rm W} \left[ \frac{2}{\hbar}\sin(\Lambdan\hbar/2) \left(1- \frac{\hbar^2}{8}\Lambdae^2\right) + \cos(\Lambdan\hbar/2) \Lambdae \right]. \eql{partexp} 
\end{align}
Using the definition of $\mL$ in \eqr{ldef} and evaluating the derivatives in \eqr{partexp} we obtain the exact quantum propagator in the mapping variable representation,
\begin{align}
\mathcal{L} = & \frac{P}{m}\ddp{}{R} - \frac{2}{\hbar} \left[V_0(R) + V_{\rm e}(R,\bq,\bp) \right] \sin\!\left(\frac{\hbar}{2}\overleftarrow {\partial_R} \overrightarrow {\partial_P} \right) \no \\
& + \frac{1}{\hbar}\left[ \bp\T \bV_\re (R) \ora \nabla_{\bq} - \bq\T \bV_\re (R) \ora \nabla_{\bp}\right] \cos\!\left(\frac{\hbar}{2}\overleftarrow {\partial_R} \overrightarrow {\partial_P}\right)  \no \\
& + \frac{1}{4} \left[\ora \nabla_\bq\T \bV_\re(R) \ora \nabla_\bq + \ora \nabla_\bp \T \bV_\re(R) \ora \nabla_\bp\right] \sin\!\left(\frac{\hbar}{2}\overleftarrow {\partial_R} \overrightarrow {\partial_P}\right), \eql{fulll}
\end{align}
one of the central results of the paper. Similar to Ref.~\citenum{kim08a}, we can define
\begin{align}
\mL = \mL_{\textrm{n}} + \mL_{\textrm{R}} + \mL_{\textrm{h}}, 
\end{align}
where
\begin{align}
 \mL_{\textrm{n}} = \frac{P}{m}\ddp{}{R} - \frac{2}{\hbar} \left[V_0(R) + V_{\rm e}(R,\bq,\bp) \right] \sin\!\left(\frac{\hbar}{2}\overleftarrow {\partial_R} \overrightarrow {\partial_P}\right) \eql{nucev}
\end{align}
corresponds to nuclear evolution on an Ehrenfest-like surface \cite{ehr27a},
\begin{align}
\mL_{\textrm{R}} = \frac{1}{\hbar}\left[ \bp\T \bV_\re(R) \ora \nabla_{\bq} - \bq\T \bV_\re(R) \ora \nabla_{\bp}\right] \cos\!\left(\frac{\hbar}{2}\overleftarrow {\partial_R} \overrightarrow {\partial_P}\right) \eql{elev}
\end{align}
corresponds to Rabi oscillations of the electronic degrees of freedom (d.o.f.) with higher-order coupling terms to nuclear motion, and
\begin{align}
\mL_{\textrm{h}} =  \frac{1}{4} \left[\ora \nabla_\bq\T \bV_\re(R) \ora \nabla_\bq + \ora \nabla_\bp\T \bV_\re(R) \ora \nabla_\bp\right] \sin\!\left(\frac{\hbar}{2}\overleftarrow {\partial_R} \overrightarrow {\partial_P}\right) \eql{difev}
\end{align}
corresponds to coupled higher-order derivatives of nuclear and electronic motion. Finally, we note that exact quantum evolution is invariant to moving the state independent potential (or any part thereof) into the nonadiabatic matrix, as shown in Appendix~\ref{ap:move}, though this will not necessarily hold when approximations are made to the propagator \cite{sto05a}.

In the following sections we analyze various analytic limits of the exact quantum propagator in \eqr{fulll}. 

\subsection{Single surface propagation}
For system on a single adiabatic surface with no coupling between nuclear and electronic d.o.f.  and all observables are in nuclear space, $V_{\rm e}(R,\bq,\bp)=0$, and \eqr{fulll} reduces to
\begin{align}
\mathcal{L}_{\rm n} = \frac{P}{m}\ddp{}{R} - V_0(R) \sin\!\left(\frac{\hbar}{2}\overleftarrow {\partial_R} \overrightarrow {\partial_P}\right), 
\end{align}
which is the conventional single-surface Moyal series propagator \cite{gro46a,hil84a,hel13a,moy49a}. 

\subsection{Electronic-only propagation}
If there are no nuclear dimensions, or no coupling between nuclear and electronic d.o.f. where the observables are in electronic space, the correlation function in \eqr{cwig} becomes
\begin{align}
c_{AB}(t) = \frac{1}{(2\pi\hbar)^{\mK}} \int d\bq \int d\bp\ [\hat A_\mS]_\rW(\bq,\bp) [\hat B(t)]_\rW(\bq,\bp) \eql{nonuc},
\end{align}
and the parts of the propagator with nuclear dependence vanish, $\mL_{\textrm{nuc}} = 0$, $\mL_{\textrm{h}} = 0$, and
\begin{align}
\mathcal{L}_{\rm R} = \frac{1}{\hbar}\left( \bp\T \bV_\re \ora \nabla_{\bq} - \bq\T \bV_\re \ora \nabla_{\bp}\right) \eql{lel}
\end{align} 
Since \eqr{lel} only contains single derivatives (i.e.\ deterministic motion) in $\bp$ and $\bq$, classical trajectories in the mapping variables will exactly reproduce the quantum correlation function in \eqr{nonuc}. To prove this, we first observe from \eqr{op} that
\begin{align}
\mathcal{L}_{\rm R} [\hat B(t)]_\rW(\bq,\bp) = & \frac{1}{2\hbar} [\mathcal{L}_{\rm R}(\bq - i\bp)\T]\mathbf{B}(t) (\bq + i \bp) + (\bq - i\bp)\T\mathbf{B}(t) [\mathcal{L}_{\rm R} (\bq + i \bp) ] \no \\
= & \frac{i}{\hbar}\frac{1}{2\hbar}[\bV_\re, (\bq - i\bp)\T\mathbf{B}(t) (\bq + i \bp)], \eql{lrb}
\end{align}
and integrating \eqr{lrb} over $t$ gives
\begin{align}
[\hat B(t)]_\rW(\bq,\bp) = \frac{1}{2\hbar} \tr[(\mathbf{C} - \hbar \mathbf{1}) e^{+i\bV_\re t/\hbar}\mathbf{B}e^{-i\bV_\re t/\hbar}]. \eql{bt}
\end{align}
Inserting \eqr{bt} into \eqr{nonuc} yields
\begin{align}
c_{AB}(t) = & \frac{1}{(\pi\hbar)^{\mK}} \frac{1}{\hbar^2} \int d\bq \int d\bp \ e^{-G/\hbar} \no\\
& \qquad \times  \tr[(\mathbf{C} - \frac{\hbar}{2}\mathbf{1}) \mathbf{A}] 
 \tr[(\mathbf{C} - \hbar \mathbf{1}) e^{+i\bV_\re t/\hbar}\mathbf{B}e^{-i\bV_\re t/\hbar}],
\end{align}
and integrating out the electronic d.o.f.\ (noting that only even powers of $q_n$ or $p_n$ survive), we find 
\begin{align}
c_{AB}(t) = \tr[\mathbf{A} e^{+i\bV_\re t/\hbar}\mathbf{B}e^{-i\bV_\re t/\hbar} ],
\end{align}
which is the conventional correlation function in the matrix representation of quantum mechanics, where $e^{\pm i \bV_\re t/\hbar}$ corresponds to the Rabi oscillations.

This analysis shows that 
\begin{align}
\bq(t) + i\bp(t) = e^{-i\bV_\re t/\hbar} [\bq(0) + i\bp(0)], \eql{matev}
\end{align}
which suggests that $q_n$ and $p_n$ can be considered the real and imaginary part respectively of the amplitude of $n$th electronic state, as suggested by the original action-angle interpretation of mapping variables \cite{mey79a,mey79b}. We caution against taking this analogy too far, since the sum of the square magnitude of amplitudes is unity, whereas the corresponding quantity in mapping variables, [$\bq\cdot\bq+\bp\cdot\bp$ in \eqr{oeval2}] has a Gaussian distribution. 


\subsection{Approximate evolution}
Truncating the exact propagator [$\mL$ in \eqr{fulll}] to different orders in $\hbar$ we find different semiclassical and quasiclassical methods emerge.

Although these methods have been very successful at investigating non-adiabatic systems, \cite{huo11a,huo12a,huo13a,kim08a,kap06a,hsi12a,hsi13a,bon05a,bon05b,ana07a} and provide ways to systematically improve the dynamics \cite{kry09a}, truncation to finite powers in $\hbar$ does not generally mean that the error in the overall correlation function scales as $\mathcal{O}(\hbar)$. \cite{hel76a} In addition, the dynamics does not normally conserve the quantum Boltzmann distribution, which can lead to spurious effects in numerical simulations \cite{hab09a}. Nevertheless, for a single electronic surface, semiclassical methods have recently been developed whereby classical trajectories conserve the quantum Boltzmann distribution \cite{hel16a}.

We firstly truncate the exact propagator [\eqr{fulll}] to $\mathcal{O}(\hbar^0)$ giving
\begin{align}
\mathcal{L}_0 = & \frac{P}{m}\ddp{}{R} - \left[V_0(R) + V_{\rm e}(R,\bq,\bp) \right] \overleftarrow {\partial_R} \overrightarrow {\partial_P}  \no\\
& + \frac{1}{\hbar}\left[ \bp\T \bV_\re(R) \ora \nabla_{\bq} - \bq\T \bV_\re(R) \ora \nabla_{\bp}\right], \eql{lscivr}
\end{align}
which is the linearized semiclassical propagator in the mapping variable representation \cite{sun97a}, corresponding to classical evolution under the mapping variable Hamiltonian in \eqr{mvham}. 
Inserting \eqr{lscivr} into \eqr{formsol}, we obtain the mapping variable LSC-IVR correlation function \cite{ana07a},
\begin{align}
c_{AB}(t)_{\rm LSC} 
= & \frac{1}{(2\pi\hbar)^{\mK+1}}\int dR \int dP \int d\bq \int d\bp\ [\hat A_\mS]_\rW\ [\hat B(0)]_\rW(R_t,P_t,\bq_t, \bp_t),
\end{align}
where $(R_t,P_t,\bq_t, \bp_t)$ are obtained by solving for the classical trajectories generated by $[\hat H]_\rW$ with initial 
conditions $(R,P,\bq, \bp)$ at time zero. 
Obtaining classical-like dynamics by truncating the propagator at $\hbar^0$ is no surprise \cite{hel76a,hel15a}---the advantage of deriving the semiclassical propagator by approximating the quantum propagator is explicit evaluation of the error in the evolution:
\begin{align}
\mathcal{L}-\mathcal{L}_0 = & -\frac{2}{i\hbar} \left[V_0(R) + V_{\rm e}(R,\bq,\bp) \right] \sum_{j=3, \ \mathrm{odd}}^{\infty} \left(\frac{i\hbar}{2}\overleftarrow {\partial_R} \overrightarrow {\partial_P}\right)^j \no\\
& + \frac{1}{\hbar}\left[ \bp\T \bV_\re(R) \ora \nabla_{\bq} - \bq\T \bV_\re(R) \ora \nabla_{\bp}\right] \sum_{j=2,\ \mathrm{even}}^{\infty} \left(\frac{i\hbar}{2}\overleftarrow {\partial_R} \overrightarrow {\partial_P}\right)^j \no\\
& + \frac{1}{4} \left[\ora \nabla_\bq\T \bV_\re(R) \ora \nabla_\bq + \ora \nabla_\bp\T \bV_\re(R) \ora \nabla_\bp\right] \sin\!\left(\frac{\hbar}{2}\overleftarrow {\partial_R} \overrightarrow {\partial_P}\right), \eql{l0er}
\end{align}
from which we see that all error terms are third order and higher derivatives, and (by construction) scale as $\mathcal{O}(\hbar)$ or greater. However, the appealing property of single-surface LSC-IVR being exact in the harmonic limit \cite{hel15a} does not extend to non-adiabatic systems unless there is no $R$ dependence in $\bV_\re$; even for the commonly-used spin-boson model of a two-state system bilinearly coupled to a harmonic bath \cite{ana13a,kim08a} the exact propagator will contain third order derivatives, which, if truncated to $\mathcal{O}(\hbar^0)$ as in \eqr{lscivr}, will lead to an error in the evolution corresponding to the third line of \eqr{l0er}.

Instead of truncating the entire propagator w.r.t.\ some order of $\hbar$, one could selectively linearize in the nuclear co-ordinates, but keep all terms in electronic d.o.f., giving
\begin{align}
\mathcal{L}_0' = & \frac{P}{m}\ddp{}{R} - \left[V_0(R) + V_{\rm e}(R,\bq,\bp) \right] \overleftarrow {\partial_R} \overrightarrow {\partial_P}  \no\\
& + \frac{1}{\hbar}\left[ \bp\T \bV_\re(R) \ora \nabla_{\bq} - \bq\T \bV_\re(R) \ora \nabla_{\bp}\right] \no\\
& + \frac{\hbar}{8} \left[\ora \nabla_\bq\T \bV_\re(R) \ora \nabla_\bq + \ora \nabla_\bp\T \bV_\re(R) \ora \nabla_\bp\right]\overleftarrow {\partial_R} \overrightarrow {\partial_P} \eql{mix}
\end{align}
which is the mixed-quantum classical evolution in mapping variables \cite{kim08a}, with an error term
\begin{align}
\mathcal{L}-\mathcal{L}_0' = & -\frac{2}{i\hbar} \left[V_0(R) + V_{\rm e}(R,\bq,\bp) \right] \sum_{j=3, \ \mathrm{odd}}^{\infty} \left(\frac{i\hbar}{2}\overleftarrow {\partial_R} \overrightarrow {\partial_P}\right)^j \no\\
& + \frac{1}{\hbar}\left[ \bp\T \bV_\re(R) \ora \nabla_{\bq} - \bq\T \bV_\re(R) \ora \nabla_{\bp}\right] \sum_{j=2,\ \mathrm{even}}^{\infty} \left(\frac{i\hbar}{2}\overleftarrow {\partial_R} \overrightarrow {\partial_P}\right)^j \no\\
& + \frac{1}{4i} \left[\ora \nabla_\bq\T \bV_\re(R) \ora \nabla_\bq + \ora \nabla_\bp\T \bV_\re(R) \ora \nabla_\bp\right] \sum_{j=3, \ \mathrm{odd}}^{\infty} \left(\frac{i\hbar}{2}\overleftarrow {\partial_R} \overrightarrow {\partial_P}\right)^j.
\end{align}
This will be exact for a spin-boson system, though the third order derivative in \eqr{mix} is not amenable to conventional classical trajectories \cite{kim08a}. Nevertheless, there exist some methods to capture higher-order terms in the Moyal series \cite{kry09a} including different evolution of forward and backward trajectories in electronic d.o.f. \cite{hsi12a,hsi13a}. 

\begin{figure}[tb]
\centering
\includegraphics[width=0.5\columnwidth]{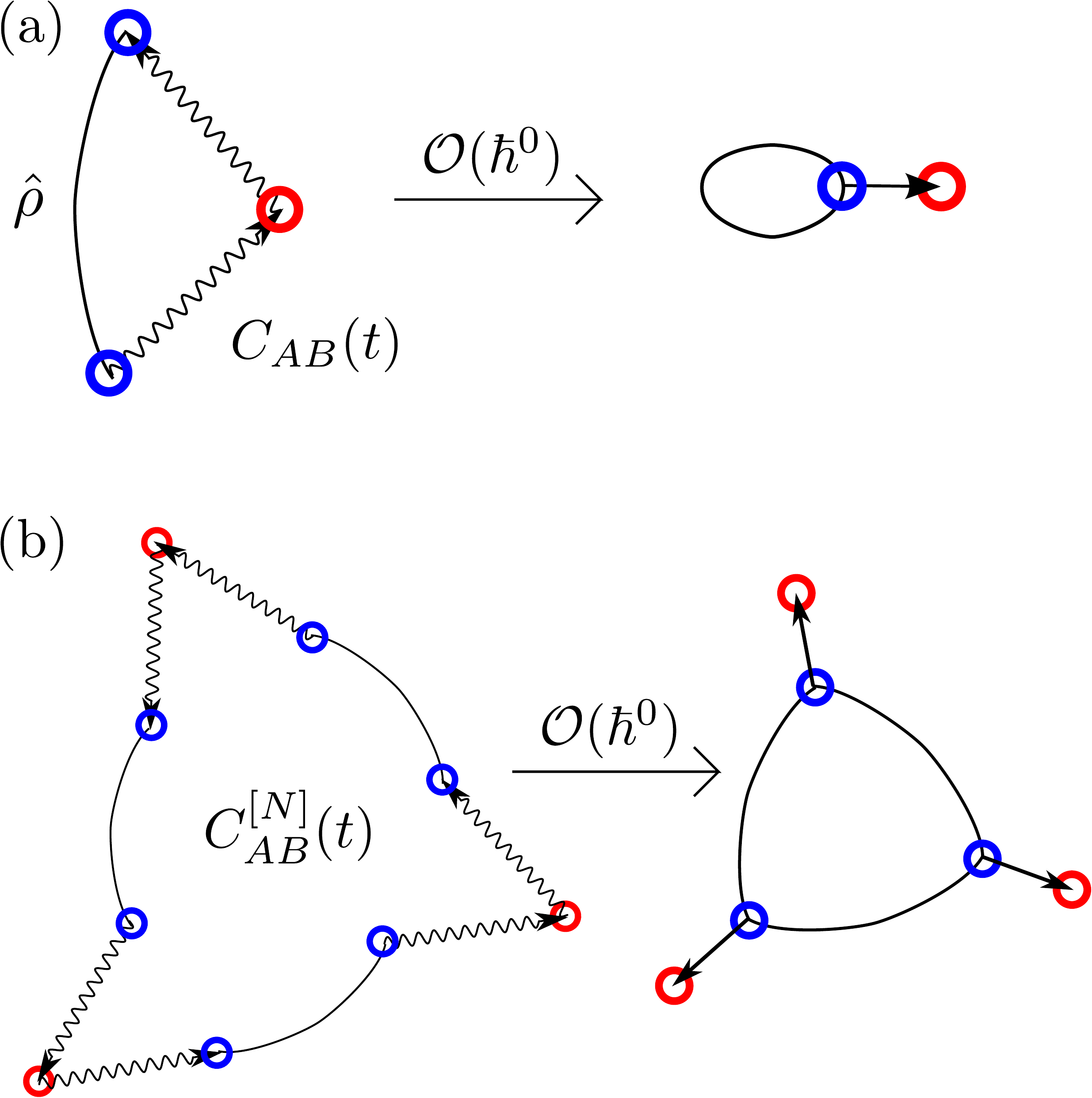}
\caption{Schematic path-integral diagrams for the correlation functions (a) $C_{AB}(t)$ and (b) $\CABNt$ in \eqr{therm} and \eqr{genk} respectively, showing the effect of truncating the real-time evolution at $\mathcal{O}(\hbar^0)$. Wavy lines represent quantum real-time evolution, curved lines imaginary time evolution (or any nonlocal density operator) and straight lines classical time-evolution. Blue and red circles represent $\hat A$ and $\hat B$ respectively, which are assumed to be local.} 
\figl{pidiagb}
\end{figure}

\section{Thermal correlation functions}
\label{sec:therm}
We consider and address the difficulties of multiple operators at zero time in mapping variable correlation functions \cite{sun98a,rab99a}. 
For a symmetrized thermal correlation function [illustrated in \figr{pidiagb}(a)]
\begin{align}
C_{AB}(t) = \tr\left[ \tfrac{1}{2}( \hat A' \hat \rho + \hat \rho \hat A') \etb\hat B\etf\right], \eql{therm}
\end{align}
where $\hat \rho$ is an arbitrary density matrix (often $\eb$), the Wigner transform of the two operators at zero time is \cite{gro46a}
\begin{align}
\left[\tfrac{1}{2}(\hat A' \hat \rho + \hat \rho \hat A')\right]_{\mathrm{W}} = [\hat A']_\rW \cos(\bm{\Lambda}\hbar/2) [\hat \rho]_\rW. \eql{cosexp}
\end{align}
If $\hat A'$ is only first order in positions and momenta (such as $\hat A' = \hat R$), or $\hat A'$ and $\hat \rho$ act in different d.o.f.\ (such as $\hat \rho$ being a nuclear Boltzmann distribution and $\hat A$ selecting a specific electronic state) then only the first term in the cosine expansion will survive and $[\tfrac{1}{2}(\hat A' \hat \rho + \hat \rho \hat A')]_{\mathrm{W}} = [\hat A']_\rW [\hat \rho]_\rW$.

Here, we evaluate \eqr{cosexp} for general $\hat A'$, written as its nuclear-only part $[\hat A_{\rm n}']_{\rW} $ and electronic part $[\mathbf{\hat A'}_{\rm e}]_{\rW}$ as in \eqr{nucel},
\begin{align}
[\hat A']_{\rW} = [\hat A_{\rm n}']_{\rW} + \frac{1}{2\hbar}\tr \left[(\mathbf{C} - \hbar \mathbf{I})[\mathbf{\hat A'}_{\rm e}]_{\rW} \right]. \eql{awig2}
\end{align}
Since any physical operator in the mapping variable representation only contains terms up to second order in $\bp$ and $\bq$ [c.f.\ \eqr{op}], we expand the cosine function in \eqr{cosexp} and without approximation truncate the trigonometric series in $\Lambdae$ to give,
\begin{align}
 [\hat A']_\rW \cos(\bm{\Lambda}\hbar/2) =  [\hat A']_\rW  \left[\left(1-\frac{\hbar^2}{8}\Lambdae^2 \right) \cos(\Lambdan \hbar/2) - \frac{\hbar}{2}\Lambdae \sin(\Lambdan \hbar/2)\right], \eql{cosexp2}
\end{align}
and by inserting \eqr{awig2} into \eqr{cosexp2} we obtain 
\begin{align}
[\hat A']_\rW \cos(\bm{\Lambda}\hbar/2) = & \left\{ [\hat A_{\rm n}']_{\rW} + \frac{1}{2\hbar}\tr \left[(\mathbf{C} - \hbar \mathbf{1})[\mathbf{\hat A'}_{\rm e}]_{\rW}\right] \right\}\cos(\Lambdan \hbar/2) \no\\
& - \frac{\hbar}{8}\left( \ora \nabla_\bp\T [\mathbf{\hat A'}_{\rm e}]_\rW \ora \nabla_\bp + \ora \nabla_\bq\T [\mathbf{\hat A'}_{\rm e}]_\rW \ora \nabla_\bq \right)\cos(\Lambdan \hbar/2) \no\\
& -\frac{1}{2} \left(\bp\T [\mathbf{\hat A'}_{\rm e}]_\rW  \ora \nabla_\bq - \bq\T [\mathbf{\hat A'}_{\rm e}]_\rW  \ora \nabla_\bp \right) \sin(\Lambdan \hbar/2), \eql{cosmoy}
\end{align}
the cosine analogue \cite{gro46a} of the sine Moyal series in \eqr{fulll}. The first line corresponds to the classical $(\hbar^0)$ term and higher nuclear derivatives, the second to a diffusion-like term in the electronic co-ordinates (with higher-order terms in nuclear d.o.f.) and the third line to mixed nuclear-electronic terms. The complicated form of \eqr{cosmoy} explains the previously noted difficulties of evaluating two operators at zero time in the mapping variable representation \cite{sun98a} and provides a mathematical framework to solve this problem. For example, the thermal population of the $\alpha$th state, where $\hat A'=\mathcal{\hat S}_\alpha = \frac{1}{2\hbar}(p_\alpha^2+q_\alpha^2 - \hbar)$, can be found using \eqr{cosmoy} as
\begin{align}
\left[\tfrac{1}{2}(\hat\mS_\alpha \hat \rho + \hat \rho \hat\mS_\alpha)\right]_{\mathrm{W}} = \frac{1}{4\hbar^2}(p_\alpha^2 + q_\alpha^2 - \hbar) \tr \left[(\bC - \hbar \mathbf{1}) \bm{\hat \rho} \right] - \frac{\bm{\hat \rho}_{\alpha\alpha}}{4}
\end{align}
where the $\rho_{\alpha\alpha}/4$ term arises from the higher derivatives on the second line of \eqr{cosmoy}.


\section{Generalized Kubo transformed correlation functions}
\label{sec:multin}
For a thermal correlation function to be computable by standard path-integral techniques, both the distribution and dynamics need to obtained in terms of classical-like variables. To achieve this, we construct the Generalized Kubo correlation function\cite{hel13a,alt13a,hel13b,hel14a} in mapping variables from which the quantum Boltzmann distribution and exact propagator can be obtained analytically. 

Consider the conventional Kubo-transformed correlation function\cite{kub57a}, 
\begin{align}
C_{AB}^{\rm Kubo}(t) = \frac{1}{\beta} \int d\lambda \tr \left[e^{-(\beta - \lambda)} \hat A e^{-\lambda} \etb \hat B \etf \right] \eql{kubo}
\end{align}
which is even and real like classical correlation functions, and can be related to the symmetric-split correlation function in \eqr{therm} by a simple Fourier transform relationship\cite{cra04a}. 
To rewrite \eqr{kubo} in a form where the Boltzmann operator is amenable to algebraic evaluation, we discretize the integral over $\lambda$ and insert position, SEO and $\etb\etf$ identities to give the Generalized Kubo transform illustrated in \figr{pidiagb}(b),\footnote{In \eqr{genk} $\hat A$ is placed in the imaginary time-evolution bra-ket; it could equivalently be placed within the real-time evolution\cite{hel13a}; here it is kept with the Boltzmann distribution for computational convenience.} 
\begin{align}
C_{AB}^{[N]}(t) = & \int d\bR \int d\bD \int d\bq \int d\bDelta \sum_{\mathbf{n,m}=1}^{\mK} \no\\
& \times \piNz \bk{\bq_{i-1} - \bDelta_{i-1}/2,R_{i-1} - D_{i-1}/2 }{{n_i}}\bra{{n_i}} \frac{1}{2}(\hat A \ebN + \ebN \hat A) \ket{{m_i}} \no\\
& \qquad \times \bk{{m_i}}{\bq_i + \bDelta_i/2,R_i + D_i/2 }\no\\
& \qquad \times  \bra{\bq_i + \bDelta_i/2,R_i + D_i/2 } \etb \hat B \etf \ket{\bq_i - \bDelta_i/2,R_i - D_i/2} \eql{genk}
\end{align}
where the operators have become
\begin{align}
\hat A = \frac{1}{N} \smkNz \hat A_k \eql{linear}
\end{align}
with $\hat A_k$ acting on the $k$th imaginary-time bead (and likewise for $\hat B$) \cite{hel15a}.
In \eqr{genk}, the operator $\hat A$ has been symmetrized around the quantum Boltzmann operator, and we use the shorthand
\begin{align}
\sum_{\mathbf{n,m}=1}^{\mK} \equiv \sum_{n_0=1}^{\mK} \ldots \sum_{n_{N-1}=1}^{\mK} \times \sum_{m_0=1}^{\mK} \ldots \sum_{m_{N-1}=1}^{\mK}. 
\end{align}
One can show by substituting \eqr{linear} into \eqr{genk} and integrating out identities that the Generalized Kubo correlation function is equivalent to the conventional Kubo correlation function \cite{kub57a} in the \largeN\ limit,\cite{hel15a}
\begin{align}
\lNti C_{AB}^{[N]}(t) = C_{AB}^{\rm Kubo}(t).
\end{align}
However, we use the Generalized Kubo transform to allow explicit evaluation of the quantum Boltzmann operator.

We now Wigner-transform \eqr{genk} giving
\begin{align}
C_{AB}^{[N]}(t) = & \frac{1}{(2\pi\hbar)^{(\mK+1)N}} \int d\bR \int d\bP \int d\bq \int d\bp \no \\
& \times  [\eb\hat A_\mS]_{\bar N} (\bR,\bP,\bq,\bp)\ [\hat B(t)]_{N} (\bR,\bP,\bq,\bp) \eql{wiggenk}
\end{align}
where the $\bar N$ subscript in $[\eb_\mS \hat A]_{\bar N}$ denotes that the Wigner transform links together the $i$th and $(i+1)$th bead whereas the subscript $N$ in $[\hat B(t)]_{N}$ only concern a single bead.\footnote{Of course, one could equivalently define the Generalized Kubo transform with $[\eb_\mS \hat A]_{N}$ and $[\hat B(t)]_{\bar N}$, but the former method is more algebraically convenient for determining time-evolution.} The integrals over $\bR, \bP$ and $\bD$ are $N$ dimensional whereas those over $\bq, \bp$ and $\bDelta$ are $N\times \mK$ dimensional.
In Appendix~\ref{ap:boltz} we show (dropping the $(\bR,\bP,\bq,\bp)$ dependence for clarity)
\begin{align}
[\eb_\mS \hat A]_{\bar N} = [\hat A]_{N} \cos(\bm{\Lambda}_N\hbar/2) [\eb_\mS]_{\bar N}, \eql{eba}
\end{align}
with $[\hat A]_{N}$ and $\bm{\Lambda}_N$ the multi-bead generalizations of $[\hat A]_\rW$ and $\bm{\Lambda}$, defined in \eqr{aa} and \eqr{lamgen} respectively. The quantum Boltzmann distribution in the path-integral representation is evaluated explicitly (in Appendix~\ref{ap:boltz}) as 
\begin{align}
[\eb_\mS]_{\bar N}  = & \frac{2^{(\mK+1)N}}{\hbar^N} \left(\frac{m}{2\pi\betaN\hbar^2}\right)^{N/2} \ e^{-G_N/\hbar} \no\\
& \times \int d\bD \ e^{i\smiNz P_i D_i \hbar}  e^{-m\smiNz [(R_{i-1}-R_i) - (D_{i-1}+D_i)/2]^2/\betaN\hbar^2} \no\\
& \times  \tr \left[\piNz  \mathbf{M}(R_{i-1} - D_{i-1}/2) \mathbf{M}(R_i + D_i/2) \left(\mathbf{C}_i - \frac{\hbar}{2}\mathbf{1}\right) \right] \eql{napot}
\end{align}
where $G_N$ and $\bC_i$ are the multi-bead generalizations of $G$ and $\bC$ [defined in \eqr{gndef} and \eqr{cidef}] and $\mathbf{M}(R) = e^{-\betaN \bV_\re(R)/2}$.
%
The Wigner-transformed real-time evolution is given as
\begin{align}
[\hat B(t)]_{N} = & \int d\bD \int d\bDelta \piNz e^{iP_i D_i \hbar} e^{i \bp_i \cdot \bDelta_i/\hbar} \no\\
& \times \bra{\bq_i - \bDelta_i/2,R_i - D_i/2 } \etb \hat B \etf \ket{\bq_i + \bDelta_i/2,R_i + D_i/2} \eql{bbig}.
\end{align}



\subsection{Generalized Kubo propagator}
In order to determine the Generalized Kubo propagator, we differentiate $\CABNt$ in  \eqr{difcor} with respect to $t$,
\begin{align}
\dd{}{t} C_{AB}^{[N]}(t) = & \frac{1}{(2\pi\hbar)^{(\mK+1)N}} \int d\bR \int d\bP \int d\bq \int d\bp \ [\eb_\mS \hat A]_{\bar N} \left[\frac{i}{\hbar}[\hat H, \hat B(t)]\right]_{N} 
\end{align}
where
\begin{align}
\left[\frac{i}{\hbar}[\hat H, \hat B(t)]\right]&_{N} \no\\
= & \int d\bD \int d\bDelta \left[\piNz e^{iP_i D_i \hbar} e^{i \bp_i \cdot \bDelta_i/\hbar}\right] \no\\
& \times \smjNz \frac{i}{\hbar} \bra{\bq_j - \bDelta_j/2,R_j - D_j/2 } \left[\hat H, \hat B(t) \right]\ket{\bq_j + \bDelta_j/2,R_j + D_j/2} \no\\
& \qquad \times \pinjNz \bra{\bq_i - \bDelta_i/2,R_i - D_i/2 } \hat B(t) \ket{\bq_i + \bDelta_i/2,R_i + D_i/2}. 
\end{align}
Because the commutator is in a bra-ket containing variables of a single ($j$th) bead, the Moyal series can be determined as for the one-bead case and then summed over all beads, giving
\begin{align}
\mathcal{L}^{[N]} = [\hat H_N]_\rW \sin(\bm{\Lambda}_N\hbar/2)
\end{align}
where
\begin{align}
[\hat H_N]_\rW = &\smiNz [\hat H_i]_\rW,
\end{align}
and $\bm{\Lambda}_N$ is the multi-bead form of the derivative operator defined in \eqr{lamgen}. This allows the generalized Kubo correlation function to be formally expressed in mapping variables as 
\begin{align}
C_{AB}^{[N]}(t) = & \frac{1}{(2\pi\hbar)^{(\mK+1)N}} \int d\bR \int d\bP \int d\bq \int d\bp \no\\  
& \times \left\{ [\hat A]_{N} \cos(\bm{\Lambda}_N\hbar/2) [\eb_\mS]_{\bar N} \right\} e^{\mL^{[N]} t} [\hat B(0)]_{N},
\end{align}
and calculating the full propagator explicitly, analogous to \eqr{nucev}--\eqr{difev}, gives
\begin{align}
\mathcal{L}^{[N]} = & \smiNz \Bigg\{ \frac{P_i}{m}\partial_{R_i} - \frac{2}{\hbar} \left[V_0(R_i) + V_{\rm e}(R_i,\bq_i,\bp_i) \right] \sin\!\left(\frac{\hbar}{2}\ola {\partial_{R_i}} \ora {\partial_{P_i}} \right) \no \\
& \qquad + \frac{1}{\hbar}\left[ \bp_i \bV_\re(R_i) \ora \nabla_{\bq_i} - \bq_i \bV_\re(R_i) \ora \nabla_{\bp_i}\right] \cos\!\left(\frac{\hbar}{2} \ola {\partial_{R_i}} \ora {\partial_{P_i}} \right) \no \\
& \qquad + \frac{1}{4} \left[\ora \nabla_{\bq_i} \bV_\re(R_i) \ora \nabla_{\bq_i} + \ora \nabla_{\bp_i} \bV_\re(R_i) \ora \nabla_{\bp_i} \right] \sin\!\left(\frac{\hbar}{2}\ola {\partial_{R_i}} \ora {\partial_{P_i}} \right)\Bigg\}. \eql{ln}
\end{align}
The distribution in \eqr{napot} and this propagator is the second major result of this paper. 

The generalized Kubo Liouvillian $\mathcal{L}^{[N]}$ corresponds to the motion of $N$ individual and independent replicas of the system, connected at zero time through the quantum Boltzmann operator. Consequently, it shares many properties with the simpler Liouvillian $\mL$ in \eqr{fulll} since there are no cross terms in $\mathcal{L}^{[N]}$ between different beads. Providing no approximation is made to the evolution, the results in Appendices~\ref{ap:move} and \ref{ap:adjoint} hold and the correlation function is invariant to placing the state-independent potential in the diabatic matrix. For a single surface it reduces to the conventional Moyal series [summed over beads as in Eq.~(43) of Ref.~\citenum{hel15a}], and truncation of \eqr{ln} to $\mathcal{O}(\hbar^0)$ gives LSC-IVR in the multi-bead representation, as detailed in Appendix~\ref{ap:trunc}.

\section{Conclusions}
\label{sec:con}
In this article we have derived the exact nonadiabatic quantum propagator in the mapping variable representation and shown how its approximation leads to pre-existing approximate methods, briefly discussing the evaluation of multiple operators at zero time. Using the Generalized Kubo transform we have then obtained an analytic expression for the thermal distribution and its associated propagator.

Future research includes determining computationally tractable but accurate approximations to the exact nonadiabatic propagator that, for instance, reproduce the correct Rabi oscillations and preserve the quantum Boltzmann distribution. These could be quasiclassical or linearized models \cite{cot16a,huo11a,huo12a,kim08a,kap06a,hsi12a,hsi13a,bon05a,bon05b}, nonadiabatic generalizations of Matsubara dynamics \cite{hel15a,hel15b,hel15c,hel16a} and may lead to methods similar to nonadiabatic CMD \cite{lia02a} and RPMD \cite{ana13a,hel11a,men11a}.

\section*{Acknowledgements}
The authors gratefully acknowledge funding from a National Science Foundation EAGER grant (Award No. CHE-1546607). N.A. additionally acknowledges support from a Sloan Foundation Fellowship, and T.J.H.H. a Research Fellowship from Jesus College, University of Cambridge and comments on the manuscript from Austin T.\ Green. 

\appendix
\section{The exact propagator is invariant to including $V_0(R)$ in $\bV_\re(R)$}
\label{ap:move}
To prove that the exact evolution is invariant to placing $V_0(R)$ (or any constant w.r.t.\ $\bp$ and $\bq$) inside the electronic evolution $V_{\rm e}(R,\bq,\bp)$, we return to the correlation function in \eqr{difcor}, noting
\begin{align}
\dd{}{t} c_{AB}(t) = & \frac{1}{(2\pi\hbar)^{\mK+1}}\int dR \int dP \int d\bq \int d\bp\ [\hat A_\mS]_{\rm W}  \mathcal{L} [\hat B(t)]_{\rm W}  \no \\
= & \frac{1}{(2\pi\hbar)^{\mK+1}}\int dR \int dP \int d\bq \int d\bp\ [\hat B(t)]_{\rm W}   \mathcal{L^\dag} [\hat A_\mS]_{\rm W} 
\end{align}
where $\mathcal{L^\dag}$ is the adjoint of $\mathcal{L}$. In Appendix~\ref{ap:adjoint} we prove that, despite $\mathcal{L}$ containing derivatives up to infinite order, $\mathcal{L} = - \mathcal{L^\dag}$ as for the classical Liouvillian. By using \eqr{wigprod} in reverse, 
\begin{align}
-\mathcal{L} [\hat A_\mS]_{\rm W} = -\frac{i}{\hbar}\left[\hat H,  \sum_{n,m=1}^{\mK} \ket{n} \bra{n} \hat A \ket{m} \bra{m} \right].
\end{align}
We then define a Hamiltonian with part of the state-independent potential moved inside the nonadiabatic matrix,
\begin{align}
\hat H_{\alpha} = \frac{\hat P^2}{2m} + V_0(\hat R) - \alpha(\hat R) + \sum_{n,m=1}^{\mK} \ket{n} [V_{nm}(\hat R) + \delta_{nm} \alpha(\hat R)] \bra{m}
\end{align}
from which we observe
\begin{align}
-\frac{i}{\hbar}\left[\hat H_\alpha,  \sum_{n,m=1}^{\mK} \ket{n} \bra{n} \hat A \ket{m} \bra{m} \right] = &
  -\frac{i}{\hbar} \sum_{n,m=1}^{\mK} \ket{n} \bra{n} [\hat H,\hat A] \ket{m} \bra{m} \no\\
  = & -\frac{i}{\hbar}\left[\hat H,  \sum_{n,m=1}^{\mK} \ket{n} \bra{n} \hat A \ket{m} \bra{m} \right]
\end{align}
and therefore infer
\begin{align}
\mathcal{L}_\alpha [\hat A_\mS]_{\rm W}  = \mathcal{L} [\hat A_\mS]_{\rm W} 
\end{align}
as required. The above proof will not hold if the Moyal expansions in the propagator are truncated, nor if SEO eigenstates are absent from $[\hat A_\mS]_{\rm W} (R,P,\bq,\bp)$.

\section{Adjoint of Moyal Series Liouvillian}
\label{ap:adjoint}
Here we prove that
\begin{align}
\dd{}{t} c_{AB}(t) = & \frac{1}{(2\pi\hbar)^{\mK+1}}\int dR \int dP \int d\bq \int d\bp\ [\hat A_\mS]_{\rm W}  \mathcal{L} [\hat B(t)]_{\rm W}  \no \\
= & -\frac{1}{(2\pi\hbar)^{\mK+1}}\int dR \int dP \int d\bq \int d\bp\ [\hat B(t)]_{\rm W} \mathcal{L} [\hat A_\mS]_{\rm W}.  \eql{toprov}
\end{align}
We observe that the Liouvillian $\mathcal{L}$ defined in \eqr{ldef} can be rewritten as
\begin{align}
\mathcal{L}=\frac{i}{\hbar} [\hat H]_{\rm W}\left(e^{-i\bm{\Lambda} \hbar/2}-e^{i\bm{\Lambda} \hbar/2} \right),
\end{align}
making it sufficient to prove 
\begin{align}
\int dr \int d\zeta A (B e^{-i\hbar\Lambda/2} C) = \int dr \int d\zeta C (B e^{i\hbar\Lambda/2} A) \eql{assoc}
\end{align}
where $r$ and $\zeta$ are general position and momentum co-ordinates of which $A,B$ and $C$ are general analytic functions, and we use one dimension for simplicity, a multidimensional generalization being straightforward. With these phase space variables
\begin{align}
\Lambda = \ola{\partial_\zeta}\ora{ \partial_r} - \ola{\partial_r}\ora {\partial_\zeta} \eql{lambda2}
\end{align}
and we use the shorthand $\partial_\zeta = \ddp{}{\zeta}$, likewise for $\partial_r$. The arrows denote the direction in which the derivative acts but when not specified, all derivatives act to the right. 

From \eqr{lambda2} it immediately follows for integer $j$ that
\begin{align}
A \Lambda^j B = (-1)^j B \Lambda^j A.
\end{align}
and from the definition of the exponential
\begin{align}
 e^{-i\hbar\Lambda/2} = \sum_{j=0}^{\infty} \left(\frac{-i\hbar}{2}\right)^j \frac{1}{j!} \Lambda^j \eql{defexp}
\end{align}
so if we can show \eqr{assoc} holds for each $j$th term of the exponential individually then it will hold for the sum of those terms. 

To prove this by induction, for the $j=0$ term we have the trivial result that $A (B \Lambda^0 C)= C (B \Lambda^0 A)$ by the commutativity of multiplication of scalar functions. We then assume that the $j$th term of \eqr{assoc} holds, i.e.
\begin{align}
\int dr\int d\zeta \ A(B\Lambda^jC) = (-1)^j\int dr\int d\zeta \ C (B \Lambda^j A ) \eql{jth}
\end{align}
and consider (to within multiplicative constants) the $(j+1)$th term
\begin{align}
\int dr \int d\zeta \ A B \Lambda^{j+1} C =  & \int dr \int d\zeta \ A [(B \Lambda^{j}) \ola{\partial_\zeta}\ora{ \partial_r}  C] - A [(B \Lambda^{j}) \ola{\partial_r}\ora{ \partial_\zeta}  C] \eql{first} \\
= &  \int dr \int d\zeta \ A [ \partial_\zeta(B \Lambda^{j})] (\partial_r C) - A [ \partial_r(B \Lambda^{j})](\partial_\zeta C) \\
= & \int dr \int d\zeta -[\partial_r A  \partial_\zeta(B \Lambda^{j})] C + [\partial_\zeta A  \partial_r (B \Lambda^{j})] C \eql{third} \\
= & \int dr \int d\zeta -(\partial_r A)  \partial_\zeta(B \Lambda^{j}) C - A [\partial_r\partial_\zeta(B \Lambda^{j})] C\no\\
&  + (\partial_\zeta A)  [\partial_r (B \Lambda^{j})] C +  A  [\partial_\zeta \partial_r (B \Lambda^{j})] C \eql{fourth} \\
= & \int dr \int d\zeta -(\partial_r A) (\partial_\zeta B) \Lambda^{j} C  + (\partial_\zeta A) (\partial_r B) \Lambda^{j} C
\end{align}
where we assume the surface terms vanish when integrating by parts\cite{hil84a} 
and derivatives only act within their brackets: for instance, so in \eqr{first} $\partial_\zeta$ does not act on $A$.
We now define $A' = \partial_r A$, $B'= \partial_\zeta B$ and use \eqr{jth} (since $A$ and $B$ are arbitrary functions) to show
\begin{align}
\int dr \int d\zeta A B \Lambda^{j+1} C = & (-1)^j \int dr \int d\zeta -C(\partial_\zeta B)\Lambda^{j}(\partial_r A) + C(\partial_r B) \Lambda^{j} (\partial_\zeta A) \no \\
= & (-1)^{j+1} \int dr \int d\zeta C (B \Lambda^{j+1} A) \eql{sev}
\end{align}
Combining \eqr{defexp} and \eqr{sev} gives \eqr{assoc} and therefore \eqr{toprov}, as required. We note that this proof is in the framework of the 
Wigner transforms but can also be obtained by using the properties of a quantum mechanical trace and then the formula for the Wigner transform of a product.

\section{Quantum Boltzmann distribution}
\label{ap:boltz}
We define the projected Boltzmann operator 
\begin{align}
\ebN_\mS = \hat \mS \ebN \hat \mS 
\end{align}
such that
\begin{align}
& [\eb_\mS \hat A]_{\bar N}\no \\
& = \int d\bD \int d\bDelta \ \piNz e^{iP_i D_i \hbar} e^{i \bp_i \cdot \bDelta_i/\hbar} \no\\
& \quad \times  \bra{\bq_{i-1} - \bDelta_{i-1}/2,R_{i-1} - D_{i-1}/2 } \frac{1}{2}(\hat A \ebN_\mS + \ebN_\mS \hat A)\ket{\bq_i + \bDelta_i/2,R_i + D_i/2 }. 
\end{align}
Using \eqr{wigprod}, and placing cross terms between adjacent beads in the Boltzmann operator and not in $\hat A$ gives
\begin{align}
[\eb_\mS \hat A]_{\bar N} = [\hat A]_{N} \left[ \smiNz\cos(\bm{\Lambda}_i\hbar/2)\right] [\eb_\mS]_{\bar N} \eql{ebasum}
\end{align}
where
\begin{align}
 [\hat A]_{N} =  & \int d\bD \int d\bDelta \piNz e^{iP_i D_i \hbar} e^{i \bp_i \cdot \bDelta_i/\hbar} \no\\
& \times \bra{\bq_{i} - \bDelta_{i}/2,R_{i} - D_{i}/2 }\hat A \ket{\bq_i + \bDelta_i/2,R_i + D_i/2 }.  \eql{aa}
\end{align}
For a linear operator as defined in \eqr{linear}, we can reduce \eqr{aa} to
\begin{align}
[\hat A]_{N} = & \frac{1}{N} \smkNz [\hat A_k]_\rW,
\end{align}
a sum over individual Wigner-transformed $\hat A_k$. Since there are no cross terms between beads in $[\hat A]_{N}$, the summation over derivatives in \eqr{ebasum} can be taken inside the cosine function to obtain \eqr{eba} with a generalized $\bm{\Lambda}$ operator
\begin{align}
\bm{\Lambda}_N = \smiNz \bm{\Lambda}_i \eql{lamgen}
\end{align}
where $\bm{\Lambda}_i$ is \eqr{lambdadef} acting on the $i$th path-integral bead.

We then evaluate the quantum Boltzmann distribution in terms of SEO eigenstates 
\begin{align}
[\eb_\mS]_{\bar N} = & \int d\bD \int d\bDelta  \sum_{\mathbf{n,m}=1}^{\mK}  \piNz e^{iP_i D_i \hbar} e^{i \bp_i \cdot \bDelta_i/\hbar} \bk{\bq_{i-1} - \bDelta_{i-1}/2}{{n_i}} \no\\
& \times  \bra{R_{i-1} - D_{i-1}/2,{n_i}}\ebN \ket{R_i + D_i/2, {m_i}} \bk{{m_i}}{\bq_i + \bDelta_i/2} \\
= & \left(\frac{2}{\hbar}\right)^{N} \frac{1}{(\pi\hbar)^{\mK N/2}}   \int d\bD \int d\bDelta   \left\{ \piNz e^{iP_i D_i \hbar} e^{i \bp_i \cdot \bDelta_i/\hbar} e^{-(\bq_i\cdot\bq_i + \bDelta_i\cdot\bDelta_i/4)/\hbar} \right\}\no\\
& \qquad \times \tr\left[\piNz \mathbf{K}_i (\bq_i + \bDelta_i/2) \otimes (\bq_{i} - \bDelta_{i}/2)\T\right] \eql{vecnot}
\end{align}
where we use vector notation for convenience and define the nuclear Boltzmann matrix as
\begin{align}
(\mathbf{K}_i)_{nm} = \bra{R_{i-1} - D_{i-1}/2,{n_i}}\ebN \ket{R_i + D_i/2, {m_i}}.
\end{align}
We evaluate the integrals over mapping variables in \eqr{vecnot} in a similar method to \eqr{aeval} and Ref.~\citenum{ana13a},
\begin{align}
[\eb_\mS]_{\bar N} = &  \frac{2^{(\mK+1)N}}{\hbar^N}  e^{-G_N/\hbar} \int d\bD \ e^{i\smiNz P_i D_i \hbar} \ \tr \left[\piNz \mathbf{K}_i \left(\mathbf{C}_i - \frac{\hbar}{2}\mathbf{1}\right) \right] \eql{elbit}
\end{align}
where 
\begin{align}
G_N= & \smiNz \bq_i \cdot \bq_i + \bp_i \cdot \bp_i , \eql{gndef} \\
\mathbf{C}_i = & (\bq_i + i \bp_i)\otimes (\bq_i - i \bp_i)\T. \eql{cidef}
\end{align}
Evaluation of the $\mathbf{K}_i$ matrices is more complicated than for a conventional ring polymer expression due to the presence of the `stretch' variables $\bD$. We choose to symmetrically split the quantum Boltzmann distribution (although similar results are obtained with an asymmetric splitting),
\begin{align}
\lNti \ebN = e^{-\betaN \hat V/2} e^{-\betaN \hat T} e^{-\betaN \hat V/2}
\end{align}
and since the nuclear kinetic energy operator is, by construction, diagonal in the diabatic basis,
\begin{align}
\mathbf{K}_i = \sqrt{\frac{m}{2\pi\betaN\hbar^2}} \mathbf{M}(R_{i-1} - D_{i-1}/2) e^{-m[(R_{i-1}-R_i) - (D_{i-1}+D_i)/2]^2/\betaN\hbar^2} \mathbf{M}(R_i + D_i/2) \eql{nucbit}
\end{align}
where $\mathbf{M}(R) = e^{-\betaN \bV_\re(R)/2}$ is an exponential matrix. Combining \eqr{elbit} and \eqr{nucbit} gives \eqr{napot}. 
For a general potential, the stretch $\bD$ cannot be integrated out from \eqr{napot} without approximation due to its presence in the exponential 
potential matrices $\mathbf{M}$, such that \eqr{napot} is qualitatively different from the nonadiabatic ring-polymer potential\cite{ana13a}. 
We also observe that there are no spring terms in electronic degrees of freedom.

\section{Truncation of $\mL^{[N]}$ to $\mathcal{O}(\hbar^0)$}
\label{ap:trunc}
Evaluating the $\hbar^0$ approximation to \eqr{ln} yields
\begin{align}
\mathcal{L}^{[N]}_0 = & \smiNz \Bigg\{ \frac{P_i}{m}\partial_{R_i} -  \left[V_0(R_i) + V_{\rm e}(R_i,\bq_i,\bp_i) \right] \frac{\hbar}{2}\ola {\partial_{R_i}} \ora {\partial_{P_i}} \no\\
& \qquad + \frac{1}{\hbar}\left[ \bp_i\T \bV_\re(R_i) \ora \nabla_{\bq_i} - \bq_i\T \bV_\re(R_i) \ora \nabla_{\bp_i}\right] \Bigg\} \eql{h0n} 
\end{align}
which, for linear observables, is identical to the LSC-IVR Kubo-transformed correlation function as discussed above,
and can be seen by considering individual terms in the sum over $\hat B_i$. \cite{hel15a}

The evolution of the electronic positions and momenta in \eqr{h0n} is identical to that used by Richardson and Thoss \cite{ric13a}, and does not in general conserve the quantum Boltzmann distribution.

\bibliography{refbig}
\end{document}